\documentclass{iau} 
\usepackage{graphicx}

\title[IAU Symposium 330.~~The Gaia sky: version 1.0] 
{The Gaia sky: version 1.0}

\author[Anthony G.A.~Brown]   
{Anthony G.A.~Brown$^1$}

\affiliation{$^1$ Sterrewacht Leiden, Leiden University, Niels Bohrweg 2, 2333 CA, Leiden,
  Netherlands\\ email: {\tt brown@strw.leidenuniv.nl}}

\pubyear{2017}
\volume{330}  
\jname{Astrometry and Astrophysics in the Gaia Sky}
\editors{A.~Recio-Blanco, P.~de Laverny, A.G.A.~Brown \& T.~Prusti, eds.}
\begin{document}

\maketitle
\newcommand{\gaia}{\textit{Gaia}}
\newcommand{\gdr}{\gaia~DR1}
\newcommand{\gdrtwo}{\gaia~DR2}
\newcommand{\hip}{\textit{Hipparcos}}
\newcommand{\tyc}{\textit{Tycho-2}}

\begin{abstract}
  In this contribution I provide a brief summary of the contents of {\gdr}. This is followed by a
  discussion of studies in the literature that attempt to characterize the quality of the
  \textit{Tycho-Gaia} Astrometric Solution parallaxes in {\gdr}, and I point out a misconception
  about the handling of the known systematic errors in the {\gdr} parallaxes. I highlight some of
  the more unexpected uses of the {\gdr} data and close with a look ahead at the next {\gaia} data
  releases, with {\gdrtwo} coming up in April 2018.
\end{abstract}

\firstsection 
\section{Overview of \gdr}

With the announcement on September 14 2016 of the first data release from the ESA {\gaia} mission
(\gdr) the astronomical community truly entered the {\gaia} era. This data release is the
culmination of over 10 years of effort by ESA and the members of the {\gaia} Data Processing and
Analysis Consortium (DPAC). The {\gaia} satellite was launched in December 2013 to collect data that
will allow the determination of highly accurate positions, parallaxes, and proper motions for over
one billion sources brighter than magnitude $G=20.7$ in the {\gaia} white-light photometric band.
The astrometry is complemented by multi-colour photometry, measured for all sources observed by
{\gaia}, and radial velocities which are collected for stars brighter than $G_\mathrm{RVS}\sim16$ in
the pass-band of Gaia's radial velocity spectrograph. The scientific goals of the mission and the
scientific instruments on board {\gaia} are summarised in \cite[Gaia Collaboration, \etal\
(2016a)]{missionpaper}. The raw data collected during the first 14 months of the mission were
processed by the DPAC, involving some 450 astronomers and IT specialists, and turned into the first
version of the {\gaia} catalogue of the sky (\cite[Gaia Collaboration, \etal\ 2016b]{dr1paper}).

\begin{figure}[t]
  \begin{center}
    \includegraphics[height=7cm]{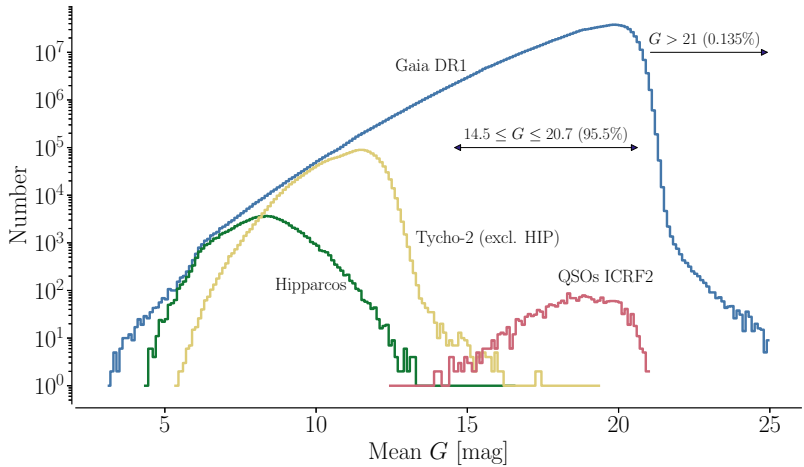}
    \caption{Magnitude distribution of the sources in the \gdr\ catalogue. The TGAS magnitude
      distribution is split into the {\hip} stars and the stars from the {\tyc} catalogue (excluding
      {\hip}). The magnitude distribution of the $\sim2000$ ICRF2 QSOs is also shown
      separately.\label{fig:gdistro}}
  \end{center}
\end{figure}

The bulk of {\gdr} consists of celestial positions $(\alpha,\delta)$ and $G$-band magnitudes for
about $1.1$ billion sources. The distribution of the {\gdr} sources in magnitude is show in
Fig.\,\ref{fig:gdistro}. With median positional accuracies of $2.3$ milli-arcsec (mas) and a spatial
resolution comparable to the Hubble Space Telescope, the {\gdr} catalogue represents the most
accurate map of the sky to date, including the most precise and homogeneous all-sky photometry,
ranging from milli-magnitude uncertainty at the bright end of the {\gaia} survey to $0.03$ magnitude
uncertainty at the faint end. In addition the combination of {\gaia} data and the positions from the
{\hip} and {\tyc} catalogues allowed the derivation of highly precise proper motions and parallaxes
for the 2 million brightest sources in \gdr\ (the so-called \textit{Tycho-Gaia} Astrometric Solution
or TGAS; \cite[Michalik \etal\ 2015, Lindegren \etal\ 2016]{tgas,agispaper}). The typical parallax
uncertainty is $0.3$~mas, while the proper motion uncertainties are about $1$~mas~yr$^{-1}$ for the
stars from the {\tyc} catalogue and as small as $0.06$~mas~yr$^{-1}$ for the stars from the {\hip}
catalogue. {\gdr} in addition contains light curves and variable type classifications for a modest
sample of some 2600 RR Lyrae and 600 Cepheid variables (\cite[Clementini \etal\ 2016]{varipipe}), as
well as the optical positions of about 2000 ICRF2 sources (\cite[Mignard \etal\ 2016]{icrf}). More
details can be found in \cite[Gaia Collaboration, \etal\ (2016b)]{dr1paper}.

\section{On the quality of the TGAS parallaxes}

Since the publication of {\gdr} various papers have treated the topic of the quality of the TGAS
parallaxes, in particular focusing on the possibility of systematic offsets with respect to
independent parallax measurements or distance estimates. The {\gdr} catalogue validation done by
DPAC \cite[(Arenou \etal\ 2017)]{validation} confirms the estimate by \cite[Lindegren \etal\
(2016)]{agispaper} that the global parallax zero-point for TGAS is at the $\pm0.1$~mas level (an
average of $\sim -0.04$~mas over various comparison samples was found, see table 2 in \cite[Arenou
\etal\ 2017]{validation}). The validation effort also confirmed the conclusion by \cite[Lindegren
\etal\ (2016)]{agispaper} that locally additional offsets at the $\pm0.2$~mas level can exist (see
for example figure 24 in \cite[Arenou \etal\ 2017]{validation}). The latter are spatially correlated
and colour-dependent. Analyses of the period-luminosity relations of local Cepheids and RR Lyrae
revealed no global offset in the TGAS parallaxes to $\sim0.02$ and $\sim0.05$~mas precision,
respectively (\cite[Casertano \etal\ 2016]{CasertanoCeps}, \cite[Sesar \etal\ 2017]{SesarRRL}).  In
contrast \cite[Jao \etal\ (2016)]{jao} and \cite[Stassun \& Torres (2016)]{stassun} find rather
large systematic offsets of $0.24$ and $0.25$ mas, where the Gaia parallaxes are too small by these
amounts compared to independently measured or predicted parallaxes for a sample of nearby stars and
a sample of eclipsing binaries, respectively.

\cite[De Ridder \etal\ (2016)]{deridder} analyzed a sample of nearby (parallaxes larger than $\sim
5$~mas) dwarfs and sub-giants as well as a sample of more distant red giants (parallaxes less than
$\sim3$~mas), for which asteroseismic distances are available based on data from the Kepler mission
(\cite[Borucki \etal\ 2010]{kepler}). In both cases the asteroseismically derived distances were
converted to parallaxes and a linear relationship,
$\varpi_\mathrm{predicted}=\alpha+\beta\varpi_\mathrm{TGAS}$, was fit between these parallaxes and
the TGAS values. For the dwarfs and sub-giants the one-to-one relation between the parallax values,
with a zero offset, was found to be a plausible model, while for the red giants \cite[De Ridder
\etal\ (2016)]{deridder} find $\alpha\sim0.3$~mas and $\beta\sim 0.75$, implying that the TGAS
parallaxes are too small. \cite[Davies \etal\ (2017)]{davies} used a sample of Red Clump (RC) stars
in the Kepler field (with parallaxes below $3$~mas) to assess the TGAS parallaxes. They predicted
the parallaxes from the apparent magnitudes of the RC stars and the assumed mean absolute magnitude
for the RC, using the $K_\mathrm{s}$ band. Fitting a linear relation they find $\alpha\sim0.24$ and
$\beta\sim1.64$, implying that the TGAS parallaxes are too large. Finally, \cite[Huber \etal\
(2017)]{huber} present an analysis of $2200$ stars in the Kepler field, from the main sequence to
the giant branch. They conclude that the offsets between the TGAS parallaxes and distances derived
from the properties of eclipsing binaries and asteroseismic samples have been overestimated for
parallaxes in the $5$--$10$~mas range and find a significantly smaller deviation than \cite[De
Ridder \etal\ (2016)]{deridder} for smaller parallaxes. They also find that the remaining
differences can be partially compensated by adopting a hotter $T_\mathrm{eff}$ scale, leaving
differences between TGAS parallaxes and asteroseismic distances at the $\sim2$ per cent level.

These apparently contradictory conclusions on the differences between TGAS parallaxes and independent
distance indicators merit a couple of remarks:
\begin{itemize}
  \item Spurious differences between two sets of parallax measurements can occur due to the effects
    of truncating the sample on the value of the parallax or that of the relative parallax
    uncertainty. For example, when comparing parallax measurements that differ in precision and
    truncating the sample on the value of the lower precision parallaxes it can be shown (assuming
    no systematic errors in either sample and Gaussian errors) that for a case similar to the
    \cite[Jao \etal\ (2016)]{jao} study (using only parallaxes larger than 40~mas in the comparison
    drawn from an underlying sample reaching to 25~mas in parallax) the mean difference in the
    parallaxes is expected to be $\sim-0.1$~mas, in the sense of high minus low precision parallax
    values. This is of the order of the offsets claimed and can be understood when considering that
    truncation on the low precision parallaxes combined with the steep increase in stars toward
    smaller true parallaxes, leads to an excess of stars in the low precision sample having
    overestimated parallaxes. It should be stressed that \cite[Jao \etal\ (2016)]{jao} are aware of
    this issue and have made the comparison for various sub-samples and also by truncating on the
    value of the high precision (TGAS) parallaxes. The differences they find cannot readily
    be explained away by the sample truncation effect only.

  \item Similarly a truncation of the sample on apparent magnitude can introduce a spurious
    difference between two sets of parallax estimates. This issue could play a role in the studies
    that use standard candles, such as RC stars, to estimate parallaxes from the apparent magnitude.
    In the study of \cite[Davies \etal\ (2017)]{davies} many of the RC stars have apparent
    magnitudes around the completeness limit of TGAS (this is also pointed out in the work by
    \cite[Gontcharov \& Mosenkov, 2017]{gontcharov}). This will lead to favouring the intrinsically
    brighter RC stars for which the parallaxes will be underestimated when using the mean RC
    absolute magnitude to calculate the parallax. This specific effect is probably not very
    important in the \cite[Davies \etal\ (2017)]{davies} study if the intrinsic spread in RC star absolute
    magnitudes is as small as derived in \cite[Hawkins \etal\ (2017)]{hawkins}. In general the
    properties of a selected sample, the properties of the parent population it is drawn from, as
    well as the survey selection functions (in both TGAS and other surveys the samples are drawn
    from) need to be well understood in order to properly interpret any differences between
    different sets of parallaxes.

  \item Offsets between TGAS and other parallax estimates that increase with the size of the
    parallax could be indicative of a scaling error in the non-astrometric parallaxes. \cite[Silva
    Aguirre \etal\ (2017)]{silva}, \cite[Huber \etal\ (2017)]{huber}, and \cite[Yildiz \etal\
    (2017)]{yildiz} point out the possibility that the $T_\mathrm{eff}$ scale used in the
    calculation of asteroseismic distances (which scale as $T_\mathrm{eff}^{2.5}$) could partly
    explain the offsets between TGAS and their distances estimates. Likewise an underestimate of the
    absolute magnitude in the case standard candles are used could explain the trend seen in the study of
    \cite[Davies \etal\ (2017)]{davies}, where the ratio of estimated to true parallax scales as
    $10^{0.2\Delta M}$, with $\Delta M$ the difference between the estimated and true absolute
    magnitude.  However $\Delta M\sim1$ is required to explain the effect found by \cite[Davies
    \etal\ (2017)]{davies}, and it is highly unlikely that the existing Red Clump absolute magnitude
    calibrations are in error by that amount. An overestimate of the extinction has a similar effect
    but again an implausibly large overestimate of the extinction toward the Kepler field would be
    required.
    
  \item Comparisons of TGAS parallaxes to independent parallax measurements or to parallaxes
    estimated by other means should always consider that systematic effects can occur in either set
    of parallaxes. Independently estimated distances that are converted to parallaxes (using
    $\varpi=1/d$) will suffer from the same non-linear transformation problems as when calculating
    distances as $1/\varpi$. Depending on the relative distance error, the error on the predicted
    parallax will be highly non-Gaussian and the mean can be biased away from the true value.

  \item The actual properties of the parallax uncertainties in both TGAS and the independent
    parallax measurements/estimates play an important role. All studies above implicitly assume
    normally distributed errors for which the values are correct. However \cite[Lindegren \etal\
    (2016)]{agispaper} show that the distribution of the normalized TGAS-Hipparcos parallax
    differences contains exponential tails, while its width hints at uncertainties being
    underestimated in one or both data sets. The differences between TGAS and previously published
    trigonometric parallaxes are modelled as Lorentzians by \cite[Jao \etal\ (2016)]{jao} in order
    to accommodate for extended wings, again hinting at partly non-Gaussian errors in the parallaxes
    of one or both samples. The parallaxes estimated from astrophysical properties of a particular
    sample of stars can likewise suffer from non-Gaussianity and/or under- or overestimation of the
    errors. Errors that deviate significantly from normal behaviour will amplify the effects
    discussed above (sample truncation, scale errors in distance estimators).

  \item A number of studies rely on stars from the Kepler field only and extrapolating the results
    to the entire TGAS catalogue is a dubious undertaking. The special validation solutions for TGAS
    discussed in appendix E of \cite[Lindegren \etal\ (2016)]{agispaper}, as well as the QSO
    analysis in \cite[Arenou \etal\ (2017)]{validation} show that there are regional systematics on
    the sky. Stars distributed over the Kepler field may well suffer from similar parallax
    systematics, as suggested by figure E.1 in \cite[Lindegren \etal\ (2016)]{agispaper}. However the
    offsets between TGAS and independent parallax estimates for the Kepler field should not blindly
    be extrapolated to other regions on the sky.
\end{itemize}

\smallskip
The studies of TGAS parallaxes carried out so far have mostly not or only partly addressed the above
issues. This makes it difficult to come to clear interpretation of the offsets seen between TGAS
parallaxes and alternative parallax measurements or estimates. I conclude that there is no strong
reason to revise the estimates of the level of systematic errors in TGAS parallaxes as presented in
\cite[Lindegren \etal\ (2016)]{agispaper} and \cite[Arenou \etal\ (2017)]{validation}.

\subsection{TGAS parallax uncertainties, treatment of (spatially correlated) systematics}

As described in \cite[Lindegren \etal\ (2016)]{agispaper} an inflation factor was applied to the
formal uncertainties on the astrometric parameters (as determined in the astrometric data
processing) to arrive at the uncertainties quoted in the {\gdr} catalogue. This inflation factor
was derived from a comparison between the TGAS and the Hipparcos parallaxes. There are indications
in various studies that the inflated parallax uncertainties may be overestimated at the 10--20\%
level (\cite[Casertano \etal\ 2016]{CasertanoCeps}, \cite[Gould \etal\ 2016]{goulderrors},
\cite[Sesar \etal\ 2017]{SesarRRL}). If desired the inflation factor applied in \cite[Lindegren
\etal\ (2016)]{agispaper} can be undone (see their section 4.1) but then this factor should be
re-estimated as part of the data analysis (see \cite[Sesar \etal\ 2017]{SesarRRL}, for an example).

Since the publication of {\gdr} there has been some confusion in the astronomical community (also
reflected in the literature) as to how to deal with the systematic uncertainties known to be present
in the TGAS astrometry, in particular for the parallaxes. This was partly caused by a misleading
statement in the paper describing {\gdr} (\cite[Gaia Collaboration \etal\ 2016b]{dr1paper}) in which
it is recommended to `consider the quoted uncertainties on the parallaxes as
$\varpi\pm\sigma_\varpi$ (random) $\pm0.3$~mas (systematic)'. This creates the impression that the
typical $0.3$ mas systematic uncertainty should be added in quadrature to the uncertainty quoted in
the {\gdr} catalogue. It should be stressed here that this {\em should not be done}. The reason is
that the calibration of the TGAS parallax uncertainties by comparison to the Hipparcos parallaxes
automatically leads to the inclusion of the local systematics in the quoted uncertainty. There is no
simple recipe to account for the systematic uncertainties. The advice is to proceed with one's
analysis of the {\gdr} data using the uncertainties quoted in the catalogue (possibly with a
modified inflation factor), but to keep the systematics in mind when interpreting the results of the
data analysis.

As illustrated in \cite[Arenou \etal\ (2017)]{validation} and \cite[Lindegren \etal\
(2016)]{agispaper} the systematic uncertainties on the parallaxes vary over the sky and are
spatially correlated in the sense that the systematics over small patches of the sky tend to be in
the same direction. No attempt was made during the {\gdr} processing and validation to derive a
correlation length scale. \cite[Zinn \etal\ (2017)]{Zinn} made use of the precise asteroseismic
distances for stars in the Kepler field to calibrate the spatial correlation length scale of
systematic parallax uncertainties in {\gdr}. They also provide a model of the spatial correlations
that can be used to construct a covariance matrix for data analyses that involve TGAS parallaxes. It
is not obvious that this finding for the Kepler field holds for the entire sky, but it does provide
a useful estimate of the local correlations and perhaps the appropriate values of the correlation
length for other sky regions can be estimated as part of the data analysis.

\section{Science from \gdr}

Notwithstanding the complexity of dealing with its error characteristics, the scientific
exploitation of the first {\gaia} data release has been taken up enthusiastically by the world-wide
astronomical community, as evidenced by the numerous workshops organized to collectively work on the
analysis of {\gaia} data\footnote{For example the `Gaia Sprints' (\texttt{http://gaia.lol/}), and the
Gaia 2016 Data Workshop (\texttt{https://www.cosmos.esa.int/web/gaia-2016-data-workshop/home}).},
and the over 300 papers that have appeared in the literature since September 14 2016 which are based
on or make use of the {\gdr} data.

The Gaia Collaboration has published two performance verification papers that provide a new
inventory of the nearby open clusters (\cite[Gaia Collaboration \etal\ 2017a]{clusterpaper}), and
a test of the TGAS parallaxes through a thorough study of the local Cepheid and RR Lyrae populations
(\cite[Gaia Collaboration \etal\ 2017b]{plpaper}), where the $K$-band period luminosity relations
show a substantial improvement in the TGAS parallaxes compared to the \textit{Hipparcos} values. I
highlight below a few of the more creative and unexpected analyses of the {\gdr} data.

\underline{\it Mapping the structure of the Magellanic clouds}. \cite[Belokurov \etal\
(2017)]{belokurov} describe a very clever method for tracking down variable stars in {\gdr}, even
though their light curves have not been published and no explicit indication is included on the
possible variability of catalogue sources (keep in mind that light curves and variable star
characterizations were included in {\gdr} only for a very modest sample of 2600 RR Lyrae and 600
Cepheid variables, see \cite[Clementini \etal\ 2016]{varipipe}). \cite[Belokurov \etal\
(2017)]{belokurov} make use of the fact that the photometric uncertainties quoted in the {\gdr}
catalogue reflect the scatter in the individual observations made for each source. This leads to
overestimates of the uncertainty on the mean $G$ band value for variable sources, making these stars
stand out in a diagram of the uncertainty in $G$ vs.\ the value of $G$. By calibrating against
samples of known variable stars \cite[Belokurov \etal\ (2017)]{belokurov} were able to identify
candidate RR Lyrae stars in a field covering the Magellanic clouds. These candidate RR Lyrae
beautifully outline the LMC and SMC and in particular reveal the bridge of old stars between the two
Milky Way companions. A combination of {\gdr} with \textit{GALEX} data (\cite[Bianchi \etal\
2014]{galex}) revealed the existence of a bridge of younger stars, offset from the bridge of old
stars and coincident with the known HI bridge between the LMC and SMC. The technique to find
variable stars in {\gdr} was also applied by \cite[Deason \etal\ (2017a)]{Deason} in order to map
the structure of the Magellanic system through Mira variables.

\underline{\it A cluster hiding behind Sirius}. The power of a high spatial resolution, high dynamic
range, all-sky star map was demonstrated nicely in the paper by \cite[Koposov \etal\ (2017)]{koposov}.
Although {\gaia} observations of bright sources suffer from CCD saturation effects, there is no need
to avoid the vicinity of even the brightest stars on the sky and hence {\gaia} can observe sources
very near such stars. \cite[Koposov \etal\ (2017)]{koposov} made use of this by creating an all-sky
map of potential source over-densities and in that way discovered a hitherto unknown star cluster
very near the brightest star in the sky, Sirius. The reality of the {\gaia}~1 cluster was confirmed
by combining the {\gdr} information with the photometry form the \textit{2MASS} (\cite[Skrutskie
\etal\ 2006]{2mass}), \textit{WISE} (\cite[Wright \etal\ 2010]{wise}), and Pan-Starrs1
(\cite[Chambers \etal\ 2016]{ps1}) surveys. \cite[Simpson \etal\ (2017)]{simpson} carried out
spectroscopic follow-up observations and concluded that {\gaia}~1 is an intermediate age
($\sim3$~Gyr) open cluster with a mass of roughly $10^4$~$M_\odot$.

\underline{\it De-noising the TGAS colour magnitude diagram}. A central goal of the {\gaia} mission
is the establishment of a precise and accurate empirical description of the colour magnitude
diagram, which opens the way to accurate luminosity calibrations of stars across the colour
magnitude diagram (CMD) and to an accurate calibration of the theoretical Hertzsprung-Russell
diagram. The measurement of stellar distances plays a fundamental role in this endeavour but
accurate distances cannot always be obtained through parallaxes alone. In particular for the more
luminous and rarer stars near the bright end of the CMD, even the end-of-mission {\gaia} parallaxes
may have relative errors above the level where one should not simply invert the parallax to obtain a
distance (see also \cite[Bailer-Jones 2015]{coryn}).  Hence it is imperative to combine multiple
pieces of information to estimate accurate distances to stars. Two papers based on {\gdr}
(\cite[Leistedt \& Hogg 2017]{leistedt} and \cite[Anderson \etal\ 2017]{anderson}) present
approaches in which the information contained in the photometry of stars (apparent brightness and
colour) is combined with the TGAS parallax information to arrive at more precise representations of
the CMD than can be obtained through TGAS parallaxes alone. In both cases a hierarchical Bayesian
model is employed albeit with a different approach to constructing the prior on the distribution of
stars in the CMD.  Both studies successfully demonstrate how this type of modelling leads to
shrinkage in the errors on the inferred distances (absolute magnitudes) of the stars, even if
strictly speaking they only provide a more precise description of the contents of the TGAS CMD,
rather than of the CMD per se (which would require folding in selection functions and considerations
on the degree to which the solar neighbourhood is representative). The hope is that eventually this
type of analysis of the {\gaia} data leads to a data-driven predictive models of stars which would
very tightly constrain our physical models of stars.

\newpage
\underline{\it The needle in the haystack}. The work by \cite[Marchetti \etal\ (2017)]{marchetti}
shows how machine learning (in this case a neural network) can be applied to large and rich data
sets such as {\gdr}. The goal of this work was to find the very few hyper-velocity stars (which were
ejected from the Galactic centre) expected to be present in the TGAS catalogue (a few hundred to a
few thousand hyper-velocity stars are expected in the full billion star {\gaia} data set). This was
done by training an artificial neural network on simulated data containing both the Milky Way and a
population of hyper-velocity stars. The optimized neural network was then applied to the TGAS data
which resulted in the identification of 80 hyper-velocity star candidates purely on the basis of
astrometric information. A careful follow-up of these candidates through the collection of radial
velocity information and the assessment of whether the orbits of the stars imply that they come from
the Galactic centre, resulted in one candidate hyper-velocity star that might be unbound from the
Milky Way and 5 candidates that appear to be bound. The results of this study greatly strengthen the
confidence that in future {\gaia} data releases (where the application of machine learning
techniques will be more important) many more hyper-velocity stars can be uncovered.

\underline{\it Stellar occultations}. Although this was not an unforeseen application of {\gdr} it is an
excellent illustration of the benefits of an accurate star map. The accurate prediction of the path
on the earth from where the occultation of a star by a minor body in the solar system can be observed
depends very much on the accuracy to which the orbit of the body is known and the accuracy to which
the star's position at the observation epoch is known. The latter is greatly improved by the
availability of {\gdr}, with more improvements expected in future {\gaia} data releases when proper
motions and parallaxes are available for all sources observed by {\gaia}. A taste of the
possibilities was provided in the summer of 2016 through the exceptional early release of the Gaia
position for a star that would be occulted by Pluto. The better knowledge of Pluto's ephemeris due
to the New Horizons flyby was combined with the more accurate {\gaia} position for the star to
enable a much more accurate prediction of the occultation path on
earth\footnote{\texttt{https://www.cosmos.esa.int/web/gaia/iow\_20160914}}. The subsequent successful
occultation campaign allowed to add a further observational point to the evolution of the
atmospheric pressure on Pluto, showing a hint that the pressure increase seen since 1988 (despite
Pluto's moving away from the Sun) is now coming to an end, perhaps indicating the start of Pluto's
predicted atmospheric `collapse' due to the lower solar flux.

The above examples are an illustration of the new and complementary ways in which astronomical
science can be pursued in the era of large surveys. Creative `playing' with the data can lead to
significant discoveries and new understanding, while at the same time the hard work of developing
statistical/numerical/data-driven methods that can efficiently deal with the large amount of
information to be uncovered is indispensable.

Finally, it should be noted that {\gdr} has quickly become the standard against which other surveys
are calibrated both astrometrically and photometrically. An example of {\gdr} serving as the
astrometric standard for another large survey is provided by SMASH (\cite[Nidever \etal\
2017]{smash}) for which the astrometry was re-reduced to the {\gdr} reference frame. A number of
proper motion catalogues have been constructed from the {\gdr} positions in combination with other
surveys.  The `Hot Stuff for One Year' proper motion catalogue (\cite[Altmann \etal\ 2017]{hsoy})
combines {\gdr} with the PPMXL (\cite[R\"oser \etal\ 2010]{ppmxl}) positions in order to derive
proper motions for over 500 million stars. The combination of Pan-Starrs1 and {\gdr} led to the GPS1
proper motion catalogue, covering three quarters of the sky (\cite[Tian \etal\ 2017]{gps1}), and
\cite[Deason \etal\ (2017b)]{gaiasdss} make use of a proper motion catalogue derived by combining
{\gdr} and SDSS (\cite[York \etal\ 2000]{sdss}). Finally the UCAC series of proper motion catalogues
was extended with the creation of UCAC5 by re-reducing the existing UCAC observations to the {\gdr}
reference frame and then combining them with the {\gdr} positions to derive new proper motions
(\cite[Zacharias \etal\ 2017]{ucac5}).

\section{Looking ahead}

Although the exploitation of the {\gdr} data is still in full swing the next data release will
arrive soon, in April 2018. {\gdrtwo} will be based on 22 months of input data and allow for a
{\gaia} stand-alone astrometric solution (so the \textit{Hipparcos/Tycho-2} positions will no longer
be used), including parallaxes and proper motions for a much larger number (of order one billion)
sources. The larger amount of data, the improvements in the various instrument calibrations, and the
introduction of colour terms in the astrometric solution will lead to large reductions of the
astrometric uncertainties. A major difference between {\gdrtwo} and {\gdr} will be the presence of
radial velocities for stars brighter than $G_\mathrm{RVS}=12$ and the availability of a broad-band
colour, $(G_\mathrm{BP}-G_\mathrm{RP})$, for all stars on an all-sky homogeneous photometric system.
Perhaps these two elements represent the biggest advance from {\gdr} to {\gdrtwo}. In addition for
stars brighter than $G=17$ the effective temperature and extinction will be determined from the
broad-band photometry in combination with the parallaxes, and a major extension of the variable star
catalogue is foreseen, including an all-sky RR Lyrae survey. Finally, the epoch astrometry for a
pre-selected list of about $10\,000$ asteroids will be released.

In connection with {\gdrtwo} it is important to be aware of the following issue concerning the
traceability of sources from {\gdr} to {\gdrtwo}\footnote{The text is a shorter version of
  \texttt{https://www.cosmos.esa.int/web/gaia/news\_20170203}}. The data processing leading up to a
  data release starts with a process that groups individual Gaia observations and links them to
  sources on the sky.  This leads to a working catalogue of sources (‘the source list’) and their
  corresponding observations, which forms the basis for the subsequent data processing. The
  algorithm that carries out the grouping and linking had been much improved before the start of the
  {\gdrtwo} processing and this led to many changes in these groups. 

When using the  Gaia data one should thus be aware that the source list for {\gdrtwo} should be
treated as independent from {\gdr}. Although the majority of sources in {\gdr} can be identified
with the same source in {\gdrtwo} through the Gaia source identifier, the improved source list will
lead to the following changes in the linking of the observations to the source identifiers for a
substantial fraction of entries in the source list:
\begin{itemize}
  \item The merging of groups of observations previously linked to more than one source will lead to
    a new source associated to the merged observations (with a new source identifier) and the
    disappearance of the original sources (along with their source identifiers).
  \item The splitting of groups of observations previously linked to one source will lead to new
    sources associated to the split groups of observations (with new source identifiers) and the
    disappearance of the original source (along with its source identifier).
  \item The list of observations linked to a source may change (and hence the source characteristics
    may change), while the source identifier remains the same.
\end{itemize}
A means to trace sources from {\gdr} to {\gdrtwo} will be provided, but a one-to-one relation will
not exist for all sources. It will then be up to the catalogue user to judge which {\gdrtwo}
source (best) matches a given {\gdr} source.

Beyond {\gdrtwo} we can look forward to {\gaia}~DR3, targeted for mid to late 2020, and {\gaia}~DR4,
targeted for the end of 2022. Details on the contents of these releases can be found on the {\gaia}
web pages\footnote{\texttt{https://www.cosmos.esa.int/web/gaia/release}}. Note that {\gaia}~DR4 will
be the final release for the nominal (5 year) {\gaia} mission. Should the {\gaia} mission be
extended, at least one additional data release is foreseen at the end of the extended mission
operations. There is much more to come!

\acknowledgements{The early release of Gaia data and the scientific success of this IAU symposium
have only been possible due to the excellence and the tireless efforts of the DPAC and ESA teams.}


\begin{thebibliography}{}
  \bibitem[Altmann \etal\ (2017)]{hsoy}
    {Altmann, M., Roeser, S., Demleitner, M., Bastian, U., Schilbach, E.}, 2017, \textit{A\&A} 600
    L4

  \bibitem[Anderson \etal\ (2017)]{anderson}
    {Anderson, L., Hogg, D.W., Leistedt, B., Price-Whelan, A.M., Bovy, J.}, 2017, arXiv:1706.0505

  \bibitem[Arenou \etal\ (2017)]{validation} {Arenou, F., Luri, X., Babusiaux, C., Fabricius, C.,
    Helmi, A., \etal}, 2017, \textit{A\&A} 599, A50

  \bibitem[Bailer-Jones (2015)]{coryn}
    {Bailer-Jones, C.A.L.}, 2015, \textit{PASP} 127, 994

  \bibitem[Belokurov \etal\ (2017)]{belokurov}
    {Belokurov, V., Erkal, D., Deason, A.J., Koposov, S.E., De Angeli, F., \etal}, 2017,
    \textit{MNRAS} 466, 4711

  \bibitem[Bianchi \etal\ (2014)]{galex}
    {Bianchi, L., Conti, A., Shiao, B.}, 2014, \textit{Advances in Space Research} Vol.\ 53, 900

  \bibitem[Borucki \etal\ (2010)]{Kepler}
    {Borucki, W.J., Koch, D., Basri, G., Batalha, N., Brown, T., \etal}, 2010, \textit{Science} 327,
    977

  \bibitem[Casertano \etal\ (2016)]{CasertanoCeps}
    {Casertano, S., Riess, A.G., Bucciarelli, B., Lattanzi, M.}, 2016, \textit{A\&A} 599, A67

  \bibitem[Chambers \etal\ (2016)]{ps1}
    {Chambers, K.C., Magnier, E.A., Metcalfe, N., Flewelling, H.A., Huber, M.E., \etal}, 2016,
    arXiv:1612.05560

  \bibitem[Davies \etal\ (2017)]{davies}
    {Davies, G.R., Lund, M.N., Miglio, A., Elsworth, Y., Kuszlewicz, J.S., \etal}, 2017,
    \textit{A\&A} 598, L4

  \bibitem[Deason \etal\ (2017a)]{Deason}
    {Deason, A.J., Belokurov, V., Erkal, D., Deason, A.J., Koposov, S.E., Mackey, D.}, 2017,
    \textit{MNRAS} 467, 2636

  \bibitem[Deason \etal\ (2017b)]{gaiasdss}
    {Deason, Alis J., Belokurov, V., Koposov, S.E., G\'omez, F.A., Grand, R.J., \etal}, 2017,
    \textit{MNRAS} 470, 1259

  \bibitem[De Ridder \etal\ (2016)]{deridder}
    {De Ridder, J., Molenberghs, G., Aerts, C., Eyer, L.}, 2016, \textit{A\&A} 595, L3

  \bibitem[Clementini \etal\ (2016)]{varipipe}
    {Clementini, G., Ripepi, V., Leccia, S., Mowlavi, N., {Lecoeur-Taibi}, I., \etal}, 2016,
    \textit{A\&A} 595, A133

  \bibitem[Gaia Collaboration \etal\ (2016a)]{missionpaper}
    {Gaia Collaboration, Prusti, T., de Bruijne, J.H.J., Brown, A.G.A., Vallenari, A., Babusiaux,
    C., \etal} 2016a, \textit{A\&A} 595, A1

  \bibitem[Gaia Collaboration \etal\ (2016b)]{dr1paper}
    {Gaia Collaboration, Brown, A.G.A., Vallenari, A., Prusti, T., de Bruijne, J.H.J., Mignard,
    F., \etal} 2016b, \textit{A\&A} 595, A2

  \bibitem[Gaia Collaboration \etal\ (2017a)]{clusterpaper}
    {Gaia Collaboration, van Leeuwen, F., Vallenari, A., Jordi, C., Lindegren, L., Bastian, U.,
    \etal}, 2017a, \textit{A\&A} 601, A19

  \bibitem[Gaia Collaboration \etal\ (2017b)]{plpaper}
    {Gaia Collaboration, Clementini, G., Eyer, L., Ripepi, V., Marconi, M., Muraveva, T., \etal},
    2017b, \textit{A\&A} in press, arXiv:1705.00688

  \bibitem[Gontcharov \& Mosenkov (2017)]{gontcharov}
  {Gontcharov, G., Mosenkov, A.}, 2017, \textit{MNRAS} 470, L97

  \bibitem[Gould \etal\ (2016)]{goulderrors}
    {Gould, A., Kollmeier, J.A., Sesar, B.}, 2016, arXiv:1609.06315

  \bibitem[Hawkins \etal\ (2017)]{hawkins}
  {Hawkins, K., Leistedt, B., Bovy, J., Hogg, D.W.}, 2017, \textit{MNRAS} 471, 722

  \bibitem[Huber \etal\ (2017)]{huber}
    {Huber, D., Zinn, J., Bojsen-Hansen, M., Pinsonneault, M., Sahlholdt, C., \etal}, 2017,
    \textit{ApJ} 844, 102

  \bibitem[Jao \etal\ (2016)]{jao}
    {Jao, W.-C., Henry, T.J., Riedel, A.R., Winters, J.G., Slatten, K.J., Gies, D.R.}, 2016,
    \textit{ApJL} 832, L18

  \bibitem[Koposov \etal\ (2017)]{koposov}
    {Koposov, S.E., Belokurov, V., Torrealba, G.}, 2017, \textit{MNRAS} 470, 2702

  \bibitem[Leistedt \& Hogg (2017)]{leistedt}
    {Leistedt, B., Hogg, D.W.}, 2017, arXiv:1703.08112

  \bibitem[Lindegren \etal\ (2016)]{agispaper}
    {Lindegren, L., Lammers, U., Bastian, U., Hern\'andez, J., Klioner, S., \etal}, 2016,
    \textit{A\&A} 595, A4

  \bibitem[Marchetti \etal\ (2017)]{marchetti}
    {Marchetti, T., Rossi, E.M., Kordopatis, G., Brown, A.G.A., Rimoldi, A., \etal}, 2017,
    \textit{MNRAS} 470, 1388

  \bibitem[Michalik \etal\ (2015)]{tgas}
    {Michalik, D., Lindegren, L., Hobbs, D.}, 2015, \textit{A\&A} 574, A115

  \bibitem[Mignard \etal\ (2016)]{icrf}
    {Mignard, F., Klioner, S., Lindegren, L., Bastian, U.,  Bombrun, A., \etal}, 2016,
    \textit{A\&A} 595, A5

  \bibitem[Nidever \etal\ (2017)]{smash}
    {Nidever, D.L., Olsen, K., Walker, A.R., Vivas, A.K., Blum, R.D., \etal}, 2017, arXiv:1701.00502

  \bibitem[R\"oser \etal\ (2010)]{ppmxl}
  {R\"oser, S., Demleitner, M., Schilbach, E.}, 2010, \textit{AJ} 139, 2440

  \bibitem[Sesar \etal\ (2017)]{SesarRRL}
    {Sesar, B., Fouesneau, M., Price-Whelan, A., Bailer-Jones, C.A.L., Gould, A., Rix, H.-W.}, 2017,
    \textit{ApJ} 838, 107

  \bibitem[Silva Aguirre \etal\ (2017)]{silva}
    {Silva Aguirre, V., Lund, M.N., Antia, H.M., Hall, W.B., Basu, S., \etal}, 2017, \textit{ApJ} 835, 173

  \bibitem[Simpson \etal\ (2017)]{simpson}
    {Simpson, J.D., De Silva, G.M., Martell, S.L., Zucker, D.B., Ferguson, A.M.N., \etal},
    2017, \textit{MNRAS} 471, 4087

  \bibitem[Skrutskie \etal\ (2006)]{2mass}
    {Skrutskie, M.F., Cutri, R.M., Stiening, R., Weinberg, M.D., Schneider, S., \etal}, 2006,
    \textit{AJ} 131, 1163

  \bibitem[Stassun \& Torres (2016)]{stassun}
    {Stassun, K.G., Torres, G.}, 2016, \textit{ApJL} 831, L6

  \bibitem[Tian \etal\ (2017)]{gps1}
  {Tian, H.-J., Gupta, P., Sesar, B., Rix, H.-W., Martin, N.F.}, 2017, \textit{ApJS} 232, 4

  \bibitem[Wright \etal\ (2010)]{wise}
    {Wright, E.L., Eisenhardt, P.R.M., Mainzer, A.K., Ressler, M.E., Cutri, R.M., \etal}, 2010,
    \textit{AJ} 140, 1868

  \bibitem[Yildiz \etal\ (2017)]{yildiz}
  {Yildiz, M., {\c C}elik Orhan, Z., \"Ortel. S., Roth, M.}, 2017, \textit{MNRAS} 470, L25

  \bibitem[York \etal\ (2000)]{sdss}
    {York, D.G., Adelman, J., Anderson, J.E., Jr., Anderson, S.F.,  Annis, J., \etal}, 2000,
    \textit{AJ} 120, 1579

  \bibitem[Zacharias \etal\ (2017)]{ucac5}
    {Zacharias, N., Finch, C., Frouard, J.}, 2017, \textit{AJ} 153, 166

  \bibitem[Zinn \etal\ (2017)]{Zinn}
  {Zinn, J.C., Huber, D., Pinsonneault, M.H., Stello, D.}, 2017, \textit{ApJ} 844, 166
\end{thebibliography}
\end{document}